\documentstyle [12pt]{article}

\begin{document}


\setlength{\baselineskip}{0.33in}

\begin{flushright}
TCD--1--93 \\
February 1993
\end{flushright}

\vspace{8mm}

\begin{center}

{\Large\bf Tachyon Splits the ($d = 2$ String)  \\
\vspace{3mm}
Black Hole Horizon and Turns it Singular}\\
\vspace{12mm}
{\large S. Kalyana Rama}

\vspace{3mm}
Mathematics Department, Trinity College, Dublin 2, Ireland. \\
email : kalyan@maths.tcd.ie \\
\end{center}

\vspace{4mm}

\vspace{4mm}

\begin{quote}
ABSTRACT.  We present a static solution
for $d = 2$ critical string theory
including the tachyon $T$ but not its potential $V(T)$.
This solution thus incorporates tachyon back reaction and,
when $T = 0$, reduces to the black hole solution.
When $T \neq 0$ one finds that (1) the Schwarzschild horizon of the
above black hole splits into two, resembling Reissner-Nordstrom
horizons and (2) the curvature scalar develops new singularities
at the horizons. These features, as we argue, will persist even
with $V(T)$ present. Some possible methods
for removing these singularities are discussed.

\end{quote}

\newpage

It is a challenging task to resolve the puzzles of
gravitating systems such as the nature of singularities, the end effect
of Hawking radiation, and information lost(?) inside the black hole.
Recently there has been a renewed effort
\cite{CGHS,review} to solve these problems
in the simpler context of two dimensional $(d = 2)$ systems, using the
string inspired toy models for quantum gravity the inspiration having
come from the discovery of black holes in  $d = 2$ critical
strings \cite{W,M,rest}.

The $d = 2$ black hole for graviton-dilaton system was discovered
as an $ \rm{SL} (2, \rm{R}) / \rm{U} (1) $ gauged
Wess-Zumino-Witten model \cite{W}; as a solution
of ${\cal O} ( \alpha' ) \; \beta$-function equations \cite{CFMP,DS}
for critical
string theory with $d = 2$ target space-time \cite{M};
and in other forms \cite{rest}.
Similar toy models in $d = 2$ space-time for quantum
gravity including matter were studied first by \cite{CGHS}
and then by many others \cite{review}
with a view of solving various puzzles of gravitating systems
in this simpler context. Most of these models have mainly
studied the graviton-dilaton system.

For a more complete story, one should also include tachyon, the only
remaining low energy degree of freedom for $d = 2$ strings (and which is
not really tachyonic for $d = 2$). Though important, its
inclusion results in non linear equations which are solved
only in a few asymptotic cases \cite{DL,review}.
The solution of $\beta$-function equations for
$d = 2$ strings including tachyons is not known.
However, solving for tachyon in the $d = 2$ string
black hole background leads to singular behaviour \cite{M}.
Thus it is important to
understand the back reaction of tachyons for $d = 2$ critical strings,
especially  because of its importance as a model for $d = 2$ quantum
gravity.

In this work we describe a static solution of the $\beta$-function
equations for the low energy $d = 2$ critical string theory
including tachyon $T$,  but not its potential $V(T)$.
This solution thus incorporates tachyon back reaction.
Though $V(T)$ is taken to be zero in obtaining the solution,
we argue that including it will not alter the qualitative
features of the solution.
The solution has, beside the ``black hole mass'' parameter,
a new parameter $\epsilon$ which is
a measure of tachyon strength. We find that: (1) The Schwarzschild horizon
of the previous black hole splits into two, resembling
Reissner-Nordstrom horizons. However, the solution cannot be
analytically continued in between the two ``horizons''.
(2) The curvature scalar $R$ has a point singularity, as for
$\epsilon = 0$, but now with $\epsilon > 0$ develops new singularities
at the horizons. Hence this solution is not really a black hole
solution in the usual sense, though for $\epsilon = 0$ it is.

The sigma model action of the $d = 2$ critical string theory for
graviton ($G_{\mu \nu}$), dilaton ($\phi$), and tachyon, in
a notation similar to that of \cite{M}, is given by
($\mu, \nu = 0, \, 1$)
\[
S_{\rm sigma} = \frac{1}{8 \pi \alpha'} \int d^2 \tilde{x} \sqrt{g}
( G_{\mu \nu} \nabla x^{\mu} \nabla x^{\nu} + \alpha' R \phi
+ 2 T )
\]
and the conformal invariance requires the following
$\beta$-function equations to be satisfied:
\begin{eqnarray}\label{beta}
R_{\mu \nu} + \nabla_{\mu} \nabla_{\nu} \phi
+ \nabla_{\mu} T \nabla_{\nu} T & = & 0 \nonumber \\
R + (\nabla \phi)^2 + 2 \nabla^2 \phi + (\nabla T)^2
+ 4 \gamma K & = & 0 \nonumber \\
\nabla^2 T + \nabla \phi \nabla T - 2 \gamma K_T & = & 0 \;
\end{eqnarray}
where $ \gamma = - \frac{2}{\alpha'} , \;
K = 1 + \frac{V}{4 \gamma}, \; V = \gamma T^2 + {\cal O} (T^3)$
and $K_T = \frac{d K}{d T}$.
These equations also follow from the target space effective action
\begin{equation}\label{target}
S = \int d^2 x \sqrt{G} \, e^{\phi} \, ( R - (\nabla \phi)^2
+ (\nabla T)^2 + 4 \gamma K ) .  \nonumber
\end{equation}
We proceed to solve the equations (\ref{beta}) in the
target space conformal gauge $ d s^2 = e^{\sigma} d u \, d v $,
with $u = x^0 + x^1$ and $v = x^0 - x^1$,
for the static case where the fields depend on $\xi \, (= u v)$
but not on $\chi \, (= u/v)$ \cite{othergg}.
Equations (\ref{beta}) then become, with
$e^{\Sigma} = \gamma \xi e^{\sigma}$ and
$\phi_n = (\xi \frac{d}{d \xi})^n \phi$, etc.\ ,
\begin{eqnarray}\label{salmon}
\Sigma_2 + \phi_1 \Sigma_1 = \Sigma_2 + \phi_2
+ T_1^2 & = & 0 \nonumber \\
T_2 + \phi_1 T_1 - \frac{1}{2} e^{\Sigma} K_T =
\phi_2 + \phi_1^2 + e^{\Sigma} K & = & 0  \; .
\end{eqnarray}
The above equations obey one Bianchi Identity \cite{CFMP}; hence, the last
equation above for example, need not be solved and can only be used to
determine some constants.
We expect the fields to evolve depending on the local metric and
hence look for their solutions in terms of $X(\xi) \equiv \Sigma_1$.
Letting $F(X) \equiv \phi_1$ and $F' = \frac{d F}{d X}$, etc.\
equations (\ref{salmon}),
using {\em e.g.}\  $F_1 = X_1 F'$, give
$\Sigma_ 2 = - X F , \; T_1^2 = X F (1+ F') $, and
\[
X F F'' + (X F' - F) (1 + F') + e^{\Sigma} T_1 K_T (X F)^{-1} = 0  .
\]
The curvature scalar is given by
$R = - 4 \gamma e^{- \Sigma} X F$.
We cannot solve these equations for the general case.
$T = 0$ will give the solutions of \cite{M} but with no
tachyon back reaction. However, for $V = 0$ these equations can be solved.
The solution incorporates tachyon back reaction and, as we argue
below, has features that are not altered when $V \neq 0$.

Thus taking $K = 1$ the equation for $F''$ gives, after dividing
it by $X F (1 + F')$,
\[
F (1 + F') = \epsilon (1 + \epsilon) X
\]
where $\epsilon (\geq 0)$ is an integration constant chosen
so that $T_1^2 \geq 0$ (equivalently
$\epsilon$ can be $\leq -1$). Taking $X = e^s$ and $F = e^s f$
the above equation can be solved to get
\begin{equation}\label{guppie}
(F - \epsilon X)^{\epsilon} \;
(F + (1 + \epsilon) X)^{(1 + \epsilon)} = {\rm constant} \; .
\end{equation}
In principle, this is  the complete solution. But it is
difficult to understand its implications. So we choose the parametrisation
\[
F - \epsilon X = l B \tau^{-1} ; \;
F + (1 + \epsilon) X = B \tau^{\delta - 1}
\]
where $\tau (\geq 0)$ is a new parameter, $l = \pm 1, \;
\delta = (1 + 2 \epsilon) (1 + \epsilon)^{-1}, \;
B = A (1 + 2 \epsilon)$ and $A$ is a constant.
Note that (i) for $l = - 1$
the constant in equation (\ref{guppie}) is not real and hence
the choice of $l$ constitutes minimal analytic continuation; and
(ii) by using residual conformal gauge transformation \cite{resgg}
we can set $B = 1$.
The above equations then give
\[
X = A \tau^{-1} (\tau^{\delta} - l); \;
F = A \tau^{-1} (\epsilon \tau^{\delta} + l (1 + \epsilon)) \; .
\]
Denoting $\dot{X} = \frac{d X}{d \tau}$, etc.\  and
noting that $\dot{X} = ((1 + \epsilon) \tau)^{-1} F$, equations
(\ref{salmon}) give
$\dot{\phi} = - X^{-1} \dot{X} , \;
\dot{T} = - \sqrt{\delta - 1} \tau^{-1} , \;
\dot{\Sigma} = - ((1 + \epsilon) \tau)^{-1} , $ and
$\tau_1 = - (1 + \epsilon) \tau X $
which can be integrated to obtain
\begin{eqnarray}\label{whale}
e^{\phi} & = & \beta_0 \tau (\tau^{\delta} - l)^{-1} \nonumber \\
T & = & - \sqrt{\delta - 1} \ln\tau   \nonumber \\
e^{\Sigma} & = & - (m l) B^2 \tau^{-(1 + \epsilon)^{-1}} \nonumber \\
\int_0^{\tau} d \tau (\tau^{\delta} - l)^{-1}
& = & A (1 + \epsilon) \ln(\frac{\tilde{\alpha}_0}{m \xi})
\end{eqnarray}
where $\beta_0$ and $\tilde{\alpha}_0$ are constants and
$m = \pm 1$. The curvature scalar is given by
\begin{equation}\label{dolphin}
R = 4 \gamma (1 + 2 \epsilon)^{-2} \tau^{- \delta}
(\tau^{\delta} - l) (\epsilon \tau^{\delta} + l (1 + \epsilon)) \; .
\end{equation}
Equations (\ref{whale}) and (\ref{dolphin}) form
the solution of equations (\ref{salmon}) with $K = 1$.
Its features are as follows.

When $\epsilon = 0$, $\tau$ can be expressed in terms of $\xi$.
The solution in (\ref{whale}) and (\ref{dolphin})
corresponds to that of \cite{M} if one takes
$A = 1, \, \beta_0 = a$, and $2 \tilde{\alpha}_0 = a \alpha'$ ($a$ is
related to black hole mass as in \cite{M}) and further
makes the branch choices,
I : $l = m = 1 , \, 1 \leq \tau \leq \infty$ and
II : $ l = m = -1 , \, 0 \leq \tau \leq \infty$ . The
respective ranges of $\xi$ are I : $\infty \geq \xi \geq 0$ and
II : $ - \tilde{\alpha}_0 \leq \xi \leq 0$. The horizon is located
at $\xi = 0$ and the curvature scalar $R$ is regular everywhere
except for a singularity at $\xi = - \tilde{\alpha}_0$.

The explicit $\tau$ integration is
not possible when $\epsilon \neq 0$ (or $\infty$).
However, one can easily
read off \cite{notefig} the following important features of the solution.

(1) From the behaviour of the integrand for $\epsilon > 0$, it can be
deduced that
$\infty \geq \xi \geq \xi_+ > 0$ for branch I and
$- \tilde{\alpha}_0 \leq \xi \leq - \xi_- < 0$ for branch II and that
$\xi_{\pm} \rightarrow 0$ as $\epsilon \rightarrow 0$
where $\xi_{\pm}$ are constants. From the
zeroes of $G_{\mu \nu}$
in Schwarzschild coordinates \cite{rtcoord} one sees that
the horizon which was
located at $\xi = 0$ for $\epsilon = 0$ now splits into two located at
$\xi = \pm \xi_{\pm}$. Thus the horizon resembles
Schwarzschild horizon for $\epsilon = 0$ and resembles
Reissner-Nordstrom horizon for $\epsilon > 0$.

(2) The curvature scalar $R$ is singular at
$\tau = 0 \, (\xi = - \tilde{\alpha}_0)$ as before, but now for
$\epsilon > 0$ it has new singularites
(with a strength proportional to $\epsilon$, for small $\epsilon$)
at $\tau = \infty \; (\xi = \pm \xi_{\pm})$. Thus the new
horizons are singular. Also, the above solutions cannot be
analytically continued into the region $- \xi_- < \xi < \xi_+$
between these two ``horizons''. (Hence the above solution is not really
a black hole solution in the usual sense, though for $\epsilon = 0$
it is).

(3) At the horizons $\xi_{\pm}$, the field $e^{\phi}$ becomes
zero. This signals a strong coupling regime as can be seen by
equation (\ref{target}).

(4) Asymptotically (branch I, $\tau \rightarrow 1_+)$
$G_{\mu \nu}$ and $\phi$ have the same behaviour as
for the case $\epsilon = 0$. Hence, the
``black hole mass'' calculated asymptotically,
using any of the methods of \cite{W,bhmass}, is of the form
$M(\epsilon) = M_0 + {\cal O}(\epsilon)$ where $M_0$ is the black
hole mass when the tachyon is zero. (However,
for the non zero tachyon case
we could not find any conserved quantity except the trivial ones given
in the equations of motion).

(5) Now consider the tachyon $T$ and its potential $V$.
It is often convenient to view
the tachyon equation in (\ref{salmon})
as an equation for  an (anti)damped oscillator where the
tachyon potential $V$ provides the restoring force and
the  couplings to graviton and dilaton provide the damping force.
Physically, when the tachyon field acquires large kinetic energy
due to gravitational interactions
the potential may be neglected. That
this is what happens can be seen by calculating the kinetic
$( (\nabla T)^2 )$ and potential $( V )$ energy terms in the
action (\ref{target})
(or, equivalently, the damping $( \phi_1 T_1 )$
and the potential $( e^{\Sigma} K_T )$ terms in the tachyon equation
in (\ref{salmon}) ). From the expressions
$ (\nabla T)^2 = 4 \gamma e^{- \sigma} T_1^2 , \,
T = - \sqrt{\delta -1} \ln\tau , \;
\phi_1 T_1 = \sqrt{\epsilon (1 + \epsilon)} X F$, and
$e^{\Sigma} K_T = e^{\Sigma} (T + {\cal O}(T^2)) $
one sees, for the solutions (\ref{whale}), that away from
the asymptotic region (branch I, $\tau >> 1$) the tachyon potential can
indeed be neglected.

However, this is not true asymptotically.
All the above terms are of the same magnitude
and tachyon potential cannot be neglected. But it is reasonable
to expect that the correct
asymptotic solution with $V$ included can be matched at some point to
(\ref{whale}) which becomes more and more valid as one nears the
horizon. That this is likely to be the case can also be seen by:
(i) taking into account the tachyon back reaction in the asymptotic
region by starting with asymptotic tachyon solution
with $V$ present and
evaluating graviton-dilaton equations including $T_1^2$ terms and
feeding the resulting values into tachyon equation again --- one finds that
the potential term $(e^{\Sigma} K_T)$ is softened;
(ii) the ${\cal O}(T^3)$ term in the tachyon potential, $V$ (see
{\em e. g.} \cite{CST}) also softens the potential
--- (i) and (ii) imply that the
tachyon potential is less steep;
and (iii) recent works  \cite{DL,review}
suggest instability due to tachyon back reaction.

Hence it seems that neglecting tachyon potential
is a reasonable approximation and that the qualitative features of
the solution presented here will persist even when $V(T)$ is properly
taken into account.

We now discuss some
possible interaction terms that may remove the singularities,
atleast the new ones:

(1) $V(T)$ : This term is unlikely to do the job as discussed above.

(2) Higher order $\alpha'$ corrections: Tseytlin \cite{tseyt} had shown
that black hole solutions of \cite{W,M, rest} survive these
corrections. Very likely, the solution given here will also survive
these corrections since the tachyon can be thought of as
an (anti)damped
oscillator gaining energy by gravitational interactions --- so that
it would have grown strong before one reaches
the region of strong curvature
where $\alpha'$ corrections are deemed important \cite{massive}.
Thus these corrections may not remove the singularities.

(3) Antisymmetric tensor, $H$ (indices on $H$ suppressed) :
This field is not there for $d = 2$ space-time. However,
for $d = 2$ toy models of a $D$ dimensional space-time, as
considered in \cite{CGHS} for example,
the resulting equations that include quantum effects
will be similar to the $\beta$-function equations.
$H$ field interactions present in such cases may possibly
remove the singularity.
Moreover, $H$ fields arise in any space-time
obtained from string theory, so it is natural
to include them.

(4) Supersymmetry : This symmetry introduces
fermions which may provide enough
repulsive force to avoid the formation of the singularities.
However, the fermions might instead form attractive condensates
and not remove the singularities.

The above (and other) possibilities are
worth pursuing. It is important to resolve this problem of new
singularities --- whether they are really generic (as it seems to be
the case) and if so how to remove them. This issue is
particularly relevent
for the string inspired toy models of $d = 2$ quantum gravity
that may answer the puzzles of $d = 4$ space-time.
The removal of the singularities seen here might
also suggest new interactions that could be important.

\vspace{2mm}

This work is supported by EOLAS Scientific Research Program \\
SC/92/206. I thank S. Sen for encouragement
and S. R. Das for pointing out the last paper in \cite{DL}.


\begin{thebibliography}{999}
\bibitem{othergg}
Equivalently one can take the static ansatz
$d s^2 = - G(x^1) (d x^0)^2 + G^{-1}(x^1) (d x^1)^2$. However, the
$\beta$-function equations for the fields
${\cal G} = \ln (4 \gamma) - \ln G , \; \psi$ (defined by
$\frac{d}{d x^1} (\psi + {\cal G} - \phi) = 0$), and $T$ turn out to
be exactly the same as the equations (\ref{salmon}) for the fields
$\Sigma, \; \phi$, and $T$ respectively.
\bibitem{resgg}
Under the transformations
$u = U^b, \, v = V^b$, which preserve the conformal gauge,
the fields are still $\chi$ independent and transform as
$(e^{\Sigma}, X, F, \phi, T)_{new} =
(b ^2 e^{\Sigma}, b X, b F, \phi, T)$.
\bibitem{notefig}
It is useful to draw a graph of $(\tau^{\delta} - l)^{-1}$
for $\epsilon = 0 , \; \infty$ and $0 < \epsilon < \infty$
for the branches I and II.
\bibitem{rtcoord}
The metric is
$d s^2 = e^{\Sigma} d t^2 - e^{- \Sigma} d r^2$
in the Schwarzschild coordinates $(r, t)$ given by
$d \chi = (2 \sqrt{- \gamma}) \chi d t$ and
$d \xi = (2 \sqrt{- \gamma}) \xi e^{- \Sigma} d r$.
The asymptotic region is $r \rightarrow \infty$
and the fields are indeed static.
\bibitem{massive}
For similar reasons, higher massive modes $A_n$ are of no help. In fact,
taking their effective action as given in \cite{CST} with zero potential,
it is easily seen that $A_n$'s have solutions similar to that of
tachyon $T$ and hence they do not remove the singularities.
\end{thebibliography}
\end{document}